\documentclass[twocolumn,twocolappendix]{aastex631}

\usepackage{graphicx}
\usepackage{amsmath}
\usepackage{subfigure}
\usepackage{amssymb}

\begin{document}
 
\title{Machine Learning Acceleration of Neutron Star Pulse Profile Modeling}
 
\author[0000-0002-8613-3140]{Preston G. Waldrop}
\affiliation{School of Physics, Georgia Institute of Technology, 837 State St NW, Atlanta, GA 30332, USA}
 
\author[0000-0003-1035-3240]{Dimitrios Psaltis}
\affiliation{School of Physics, Georgia Institute of Technology, 837 State St NW, Atlanta, GA 30332, USA}

\author[0009-0003-6348-7143]{Tong Zhao}
\affiliation{School of Physics, Georgia Institute of Technology, 837 State St NW, Atlanta, GA 30332, USA}
 
\begin{abstract}
Ray tracing algorithms that compute pulse profiles from rotating neutron stars are essential tools for constraining neutron-star properties with data from missions such as NICER. However, the high computational cost of these simulations presents a significant bottleneck for inference algorithms that require millions of evaluations, such as Markov Chain Monte Carlo  methods. In this work, we develop a residual neural network  model that accelerates this calculation by predicting the observed flux from the surface of a spinning neutron star as a function of its physical parameters and rotational phase. Leveraging GPU-parallelized evaluation, we demonstrate that our model achieves many orders-of-magnitude speedup compared to traditional ray tracing while maintaining high accuracy. We also show that the trained network can efficiently accommodate complex emission geometries, including non-circular and multiple hot spots, by integrating over localized flux predictions.
\end{abstract}
 
\keywords{Neutron stars (1108), Neural networks (1933), Algorithms (1883), Pulsars (1306)}
 

\section{Introduction} \label{sec:intro}

Machine learning has proliferated across different science and engineering disciplines. Examples in astrophysics include classification networks (e.g., to distinguish  gravitational-wave signals from uncorrelated noise; see \citealt{mly}), autoencoders (e.g., to learn the dimensionality of dynamical systems; see \citealt{autoenc}), and physics informed neural networks (e.g., to solve differential equations for fluid mechanics; see \citealt{pinns}). In this work, we focus on a different use case: using machine learning to accelerate bottlenecks in computationally expensive simulations. In particular, we focus on the use of a residual neural network to accelerate the numerical simulation of the phase-dependent flux and spectrum observed from the surface of a spinning neutron star. 

The traditional method for performing the calculation involves tracing light rays through the curved space time from an observer plane to the neutron star surface by solving the geodesic equations in the metric (\citealt{Penchenick}; \citealt{Braje}; \citealt{Cadeau}; \citealt{Psaltis2014a}). The foot points of these rays on the stellar surface are then used to calculate the flux observed for a given set of neutron star parameters and at a given rotational phase. This framework enables computational acceleration through parallelization, either across multiple CPU cores or on a GPU \citep{GRay}.

The pulse profiles generated by this method have been used to model observations from thermally emitting pulsars, infer the neutron star properties, and put constraints on the equation of state of neutron-star matter. The Neutron star Interior Composition Explorer (NICER) mission aboard the International Space Station has been performing such measurements \citep{NICER}. With data from NICER and a method to generate pulse profiles, the parameters of the neutron star are inferred via sampling methods such as Markov Chain Monte Carlos (MCMC) or Nested Sampling.

The main challenges with such inference algorithms is the large number of simulated data that need to be sampled. As a quick estimate, if we use $N_{\rm pts}$ MCMC samples and generate a marginalized 2-dimensional posterior distribution with $N_{\rm bin}$ number of bins in each dimension, the fractional Poison error in each bin will be $\epsilon\sim (N_{\rm pts}/N_{\rm bin}^2)^{-1/2}$. Achieving a fractional accuracy of 1\% in each bin, which would allow us to estimate 99-th percentile credible intervals, then requires 
\begin{equation}
N_{\rm pts}\sim 10^6 \left(\frac{N_{\rm bin}}{10}\right)^2 
\left(\frac{\epsilon}{0.01}\right)^{-2}\;.
\end{equation}
This large number of samples determines the computational resources required and the wall time it takes to achieve convergence of a sampling algorithm.

Because ray tracing algorithms are relatively slow, even with GPU parallelization, the computational bottleneck they introduce in generating pulse profiles necessitates either simplifying assumptions or a reduction in sampling resolution to keep runtimes manageable. For instance, when inferring neutron star masses and radii, it is common to adopt the simpler Schwarzschild metric over the more appropriate Hartle-Thorne metric (\citealt{Bogdanov2021}; cf.\ \citealt{Psaltis2014a}) and to limit the sampling points to a number as small as $10^4$~\citep{xpsi}. Consequently, these approximations can introduce unknown biases~\citep{xpsi}, underscoring the need for faster, more efficient methods of profile generation.

Using a residual neural network to accelerate ray tracing calculations provides an ideal solution for two reasons. First, there are already physically defined boundaries on each input parameter of the neural network. Some of these boundaries are set by the geometry of the problem, while others are set by the physics of the neutron stars (see \citealt{OzelRev2016} for physically acceptable limits on masses and radii). As a result, the neural network is not being asked to extrapolate outside of the parameter bounds that it is trained on. Second, once the neural network is trained, it is trivial to parallelize the evaluation of the network for many inputs on a GPU. This parallelization allows for: {\em (i)\/} the fast evaluation of many pulse profiles with different neutron star parameters and {\em (ii)\/} the generation of pulse profiles for neutron stars with complex emission geometries, including multiple emission regions or non-circular emission areas. The latter is possible by using the neural network to evaluate the flux observed from each infinitesimal surface element and integrating over the emission region. In this paper, we show that a neural network can indeed be trained to generate observables for neutron-star pulse profiles as a function of the input parameters with high accuracy and is substantially faster than existing ray tracing algorithms.

In \S\ref{sec:RT}, we discuss the ray tracing algorithm used to generate pulse profiles on which we trained the neural network. In \S\ref{sec:train}, we present the structure of the neural network used and, in  \S\ref{subsec:perf} and \S\ref{subsec:time}, we explore its performance in terms of accuracy and evaluation time. In \S\ref{sec:geom}, we finally demonstrate the ability of the neural network to calculate efficiently pulse profiles from pulsars with complex geometries of surface emission.

\section{Ray Tracing Algorithm} 
\label{sec:RT}

\begin{figure}[t]
    \centering
    \includegraphics[width=0.45\textwidth]{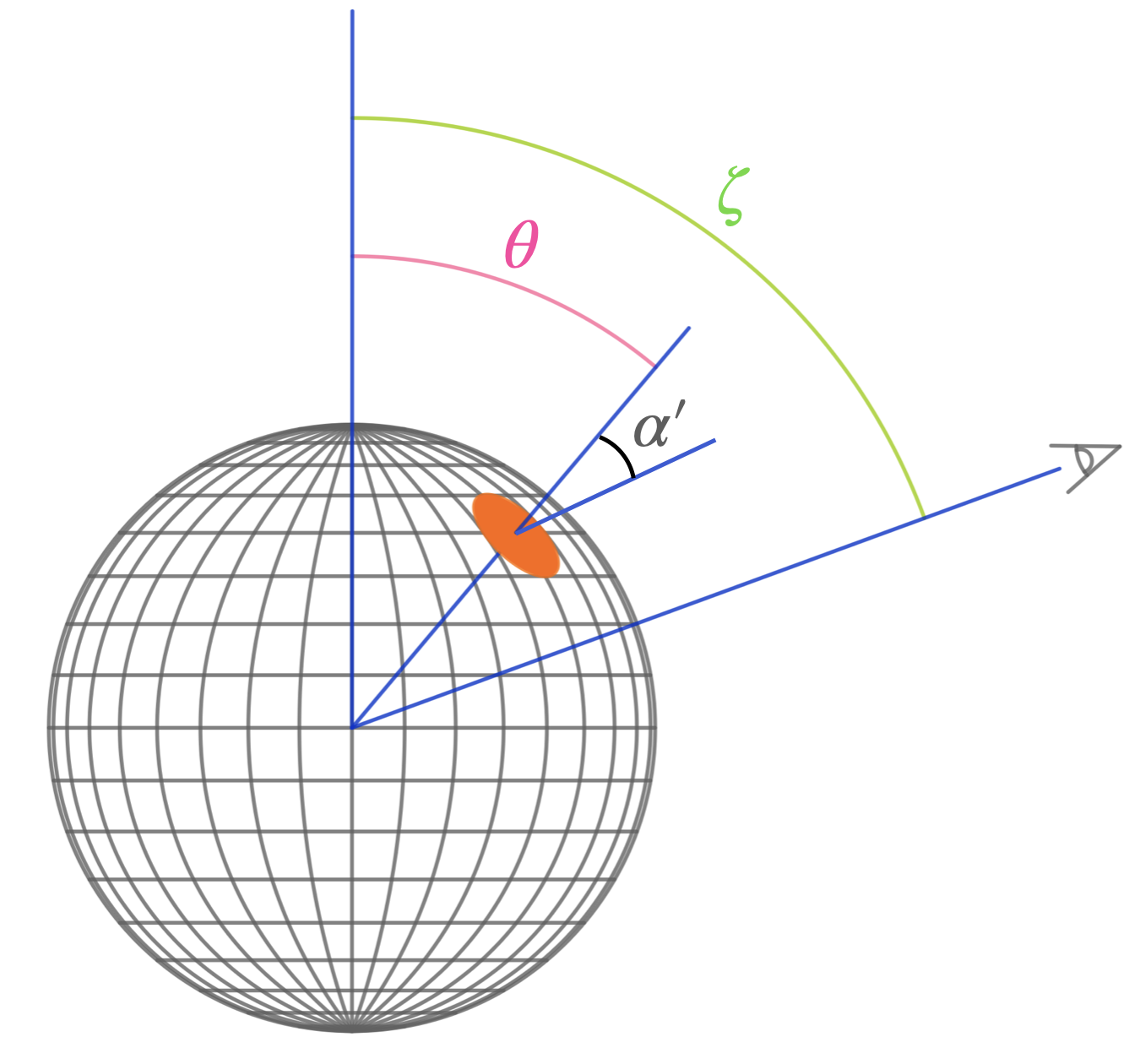}
    \caption{The relative geometry of a small surface patch on the neutron star and a distant observer; the angle $\theta$ is the colatitude of the surface patch measured with respect to the stellar spin axis, while the angle $\zeta$ measures the orientation of the observer. The azimuthal coordinate $\phi$ of the surface patch (not shown in the figure) is measured from the plane that contains the spin axis and the observer. The angle $\alpha^\prime$ with respect to the surface normal is the angle at which a photon emerging from the polar coordinates $(\theta,\phi)$ leaves the surface of the neutron star.}
    \label{fig:NSillus}
\end{figure}

In order to generate datasets on which to train the neural net, we calculate the flux from a spinning neutron star measured by an observer at infinity, for a large number of configurations of the star and of the observer. We use the ray tracing algorithm described in \citet{Baubock2012} and in \citet{Psaltis2014a}, in which null geodesics are traced backwards from the image plane of an observer at infinity to the surface of a neutron star. 

The flux measured at infinity depends on four ingredients: {\em (i)\/} the lateral temperature distribution on the neutron-star surface, {\em (ii)\/} the radial dependence of the temperature in the atmosphere, which determines the beaming of radiation emerging from the surface, {\em (iii)\/} the shape of the neutron-star surface, which sets the boundary conditions for photon propagation to infinity, and {\em (iv)} the effects of gravitational lensing caused by the gravitational field of the star through its mass, radius, and quadrupole moment. 

The lateral and radial temperature profiles determine the specific intensity $I'(E^\prime;\theta,\phi,\alpha^\prime)$ of photons at energy $E^\prime$ emerging at an angle $\alpha^\prime$ with respect to the surface normal at a location on the stellar surface with polar coordinates $(\theta,\phi)$ with respect to the rotational spin axis. Without lack of generality, we measure the azimuthal angle $\phi$ from the plane that contains the stellar spin axis and the observer. Primed quantities are measured by a local observer comoving instantaneously with the surface. 

In order to allow for any functional form of the specific intensity emerging from the stellar surface, our aim is to devise a model for the flux $dF_\infty(E_\infty)$ observed at infinity at photon energy $E_\infty$ from an infinitesimally small patch on the stellar surface. We can then calculate the total flux observed from the neutron star by integrating over all hot patches on the stellar surface. For computational accuracy, we fix the angular size of the ``small'' surface patch to $\rho=10^\circ$. As shown by \citet{spotsize}, a surface patch of this size introduces only percent-level corrections to the flux (per unit surface area of the patch) as compared to an infinitesimally small patch.

\begin{table}[t]
    \caption{Parameters Used for the Training Dataset}
    \label{tab:params}
    \begin{ruledtabular}
    \begin{tabular}{lccc}
        Parameter & Minimum & Maximum & Step \\ \hline
        Radius (km) & 9 & 14 & 0.2 \\
        Mass ($M_\odot$) & 1 & 2.4 & 0.1 \\
        Spin Frequency (Hz) & 50 & 600 & 50 \\
        $\theta$ & 15$^\circ$ & 165$^\circ$ & 7.5$^\circ$ \\
        $\zeta$ & 15$^\circ$ & 90$^\circ$ & 7.5$^\circ$ \\
    \end{tabular}
    \end{ruledtabular}
\end{table}

The external spacetime and the shape of the neutron-star surface are determined by the unknown equation-of-state of neutron-star matter and the spin frequency~$f$. For spin frequencies $f\lesssim 600$~Hz, the stellar surface can be well approximated as an oblate ellipsoid and the external spacetime is well described by the Hartle-Thorne metric~\citep{Hartle1968}. In this limit, the shape of the surface requires two parameters for its complete description: the equatorial and polar radii, $R_{\rm eq}$ and $R_{\rm p}$, respectively. The metric requires three additional parameters, the mass of the star $M$, the dimensionless spin angular momentum $a$, and the dimensionless quadrupole moment of the spacetime $q$. In order to reduce the number of model parameters, we will use the approximate relations devised in \citet{Baubock2013}, which connect the polar radius, spin angular momentum, and quadrupole moment of the star to its compactness
\begin{equation}
    u\equiv \frac{2 G M}{R_{\rm eq}c^2}
\end{equation}
and to its spin frequency and are accurate at the percent level for the conditions of interest. In order to further reduce trivial dependencies between model parameters, we will use the equatorial velocity \begin{equation}
    v_\text{eq} = \frac{2\pi f R_{\rm eq}}{c}\;.
\end{equation}

With these definitions, we use $F_\infty(E_\infty; u, v_{\rm eq}, \theta, \phi, \zeta)$ to denote the radiation flux measured at photon energy $E_\infty$ by a distant observer located at an angle $\zeta$ with respect to the spin axis of a neutron star with compactness $u$ and equatorial spin velocity $v_{\rm eq}$, emitting radiation from a small spot localized at polar coordinates $(\theta,\phi)$. Figure~\ref{fig:NSillus} depicts this geometry. 

\begin{figure}[t]
    \centering
    \includegraphics[width=0.45\textwidth]{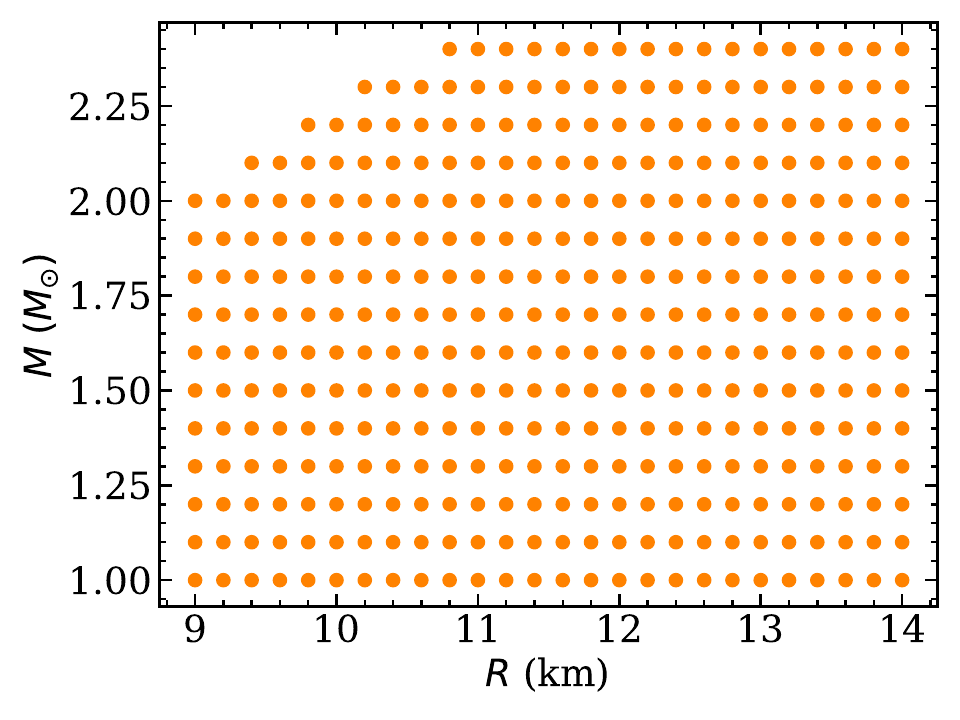}
    \caption{Neutron-star masses and radii used to generate pulse profiles with the traditional ray tracing algorithm and create the training data for the neural network.}
    \label{fig:rmparam}
\end{figure}

Finally, to remove the overall scaling of the flux with the size of the hot patch, we divide the calculated flux by the area of the patch. In geometric units, the latter is equal to 
\begin{equation}
{\cal A}=2\pi ( 1-\cos\rho)\left( \frac{Rc^2}{GM} \right)^2
\end{equation}
such that
\begin{equation}
    F = \frac{F_\infty(E_\infty; u, v_{\rm eq}, \theta, \phi, \zeta)}{2\pi ( 1-\cos\rho)\left( \frac{Rc^2}{GM} \right)^2} \;. \label{eqn:flux}
\end{equation}

\section{Generating Training Data} \label{sec:TD}

Using the ray tracing algorithm described above, we generated a dataset to train the neural network. Table~\ref{tab:params} shows the range of parameters and the step sizes used. The ranges for mass ($1M_\odot - 2.2M_\odot$) and radius (9 km - 14 km) were chosen to cover all possible values that we could expect to measure based on astrophysical considerations \citep{OzelRev2016}. We restricted the data set so that the compactness is $u<2/3$ (see Fig.~\ref{fig:rmparam}). This keeps the data out of the regime where multiple images of the surface patch are found on the image plane. 

As discussed in \S\ref{sec:RT}, for spin frequencies $\gtrsim 600$ Hz, the ellipsoid approximation for the surface of the neutron star becomes inaccurate. We have, therefore, limited the range of spin frequencies we consider to $f<600$~Hz. The ranges for the location of the surface patch ($15^\circ - 165^\circ$) and inclination of the observer ($15^\circ - 90^\circ$) were chosen so that rays were not traced around the axis of rotation, a known weakness of ray tracing algorithms in spherical polar coordinates. 

\begin{figure}[t]
    \centering
    \includegraphics[width=0.45\textwidth]{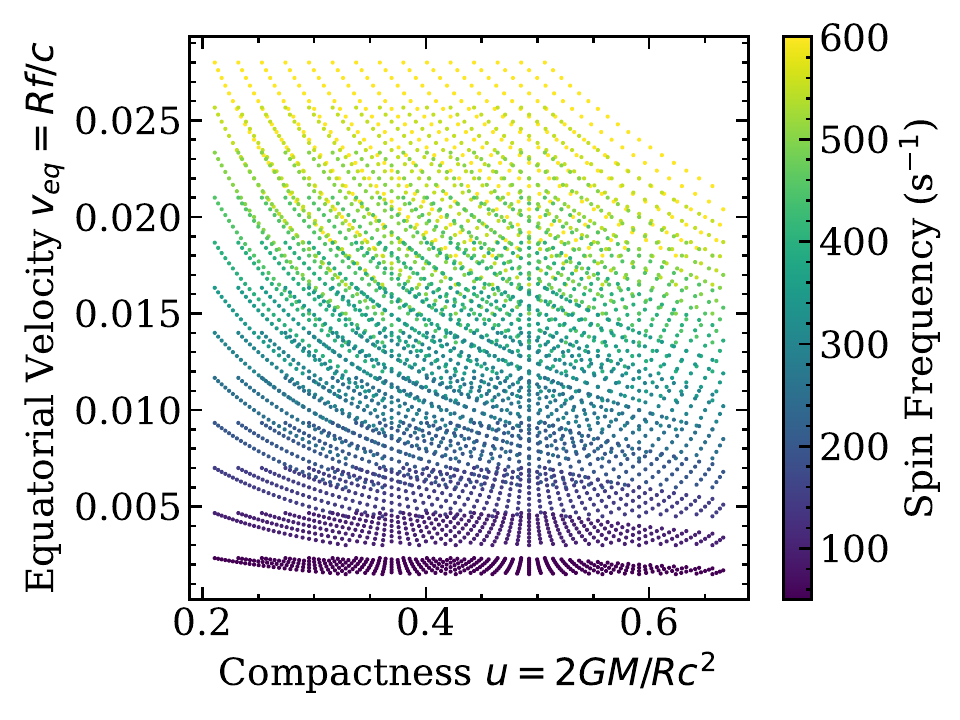}
    \caption{The parameter space used to train the neural network in the compactness vs equatorial velocity space.}
    \label{fig:uvparam}
\end{figure}

Since the metric only depends on the compactness and velocity at the equator and not on the mass, radius, and rotation rate, we train the neural network on the former set (see Figure \ref{fig:uvparam}). For each set of model parameters, we calculate the pulse profile at 127 grid points in rotational phase in [0,1). This generates a dataset of $\sim 130$ million data points. Using a standard (80\% train/20\% test) split gives $\sim 104$ million data points to train the neural network and $\sim 26$  million data points to validate the neural network. 

While the flux from an emission region on a pulsar needs to be calculated by integrating many rays, the Doppler factor, $E_\infty/E_0$, and $\cos(\alpha^\prime)$ is calculated for each ray that is traced back to the surface of the neutron star. To generate the training data for the $E_\infty/E^\prime$ and $\cos(\alpha^\prime)$ networks, we use the ray tracing algorithm to generate images of the surface of the neutron star at each radius, mass, spin frequency, and observing angle in Table \ref{tab:params}. From the ray tracing of the images, we retain $\theta$ and $\phi$ on the surface of the neutron star where the ray lands, as well as $E_\infty/E^\prime$ and $\cos(\alpha^\prime)$ for those locations. The angle $\theta$ where the ray lands on the neutron star is the same as what is used for the center of a spot relative to the axis of rotation. The angle $\phi$ from the output, that runs from $-\pi$ to $\pi$, is then converted to be the rotational phase $\varphi=\phi/2\pi$.

\section{Training of the Neural Network}\label{sec:train}

We use a fully connected Deep Neural Network. The network has six dimensionless inputs: $u$, $v_\text{eq}$, $\theta$, $\zeta$, $cos(2\pi\varphi)$, and $\sin(2\pi\varphi$), where $\varphi$ is the rotational phase running from 0 to 1. We employ $\cos(2\pi\varphi)$ and $\sin(2\pi\varphi)$ instead of just the rotational phase so that the neural network does not need to learn the fact that the pulse profiles are periodic. This has the additional benefit of telling the network that the beginning and end of the pulse profile must be continuous. 

\begin{figure}[t]
    \centering
    \includegraphics[width=0.45\textwidth]{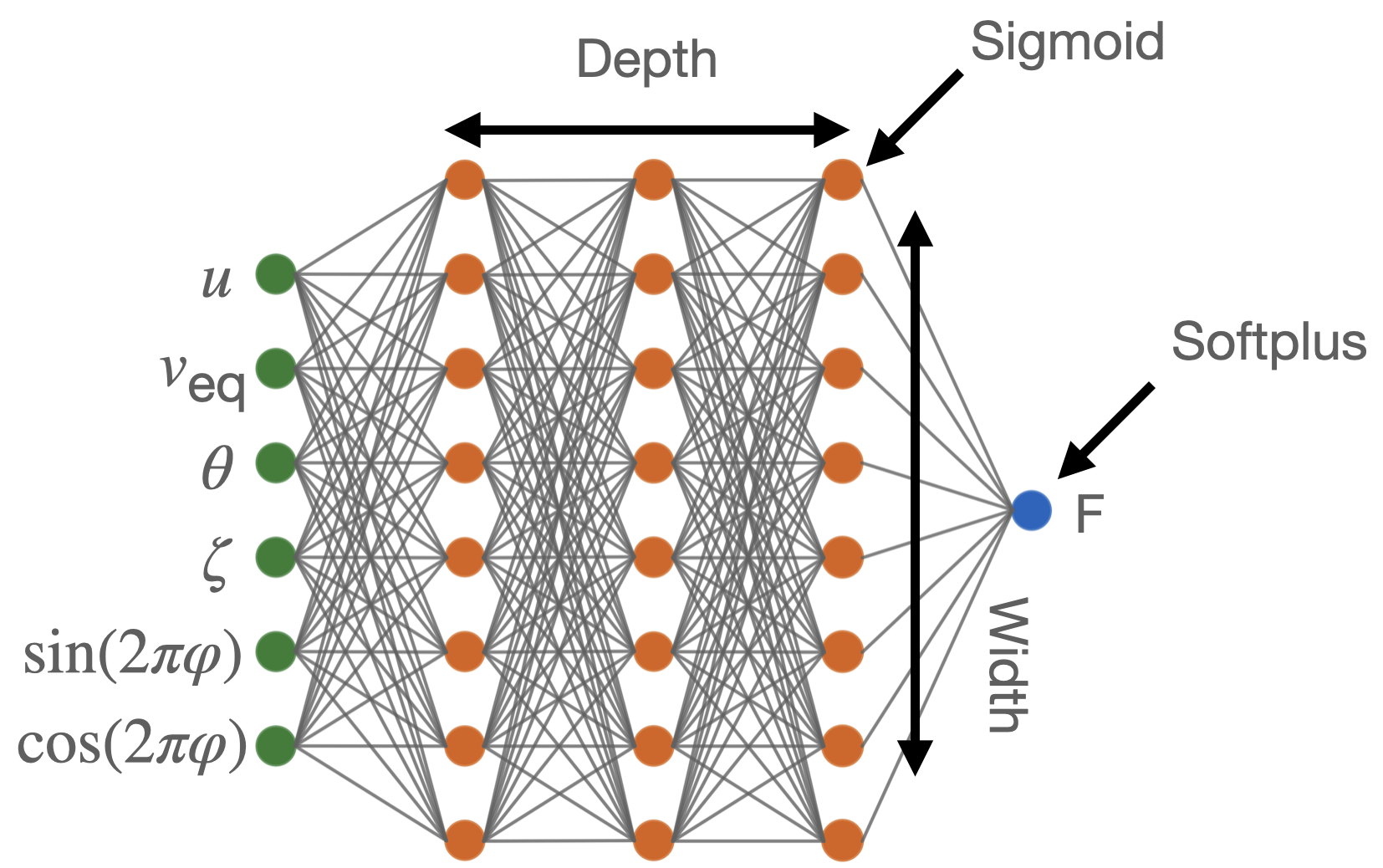}
    \caption{Illustration of the neural networks as a function of six input parameters. The input parameters are the compactness $(u)$, velocity at the equator $(v_\text{eq})$, the colatitude of the hotspot $(\theta)$, the angle of the observer from the axis of rotation $(\zeta)$, and finally the sine and cosine of the rotational phase $(\varphi)$. The output is the corresponding normalized flux. Depth is the number of hidden layers. Width is the number of neurons on each hidden layer. The specific network illustrated corresponds to a network of depth 3 and width 8.}
    \label{fig:NNillus}
\end{figure}

Using these six inputs,  we train three neural networks that output: (1) the  flux at infinity $F$, (2) the cosine of the associated beaming angle $\cos(\alpha^\prime)$, and (3) the frequency shift $g\equiv E_\infty/E^\prime$ experienced by the photons. (Note that, hereafter, we drop the prime from the beaming angle $\alpha$ for brevity). 

Figure \ref{fig:NNillus} shows an illustration of the neural network. The network is fully connected with hidden layers between the input and output. We trained neural networks with different numbers of hidden layers, i.e. depth, and different numbers of neurons on each of the hidden layers, i.e. width, to find the optimal number of trainable parameters that balances accuracy and computation time. The depths and widths tested for the best hyperparameters ranged from 2 layers to 16 layers deep in powers of 2 and the width ranged from 2 neurons per hidden layer to 1024 neurons per hidden layer in powers of 2. 

For the non-linear activation functions, we used
\begin{equation}
    \text{Sigmoid}(z) = \frac{1}{1- e^{-z}},
\end{equation}
on all hidden layers and
\begin{equation}
    \text{Softplus}(z) = \log (1+e^z),
\end{equation}
on the output layer of the network, where $z$ is the output of the previous linear layer. We used PyTorch to train the neural network with the Adam optimizer \citep{adam}, 32-bit float trainable parameters, and a mean squared error loss function,
\begin{equation}
    \text{MSELoss} = \frac{1}{N} \sum_{n=1}^{N} (x_n-y_n)^2,
\end{equation}
where $N$ is the batch size, $x_n$ is the output of the network, and $y_n$ is the target.

\subsection{Performance} \label{subsec:perf}

\begin{figure}[t]
    \centering
    \includegraphics[width=0.45\textwidth]{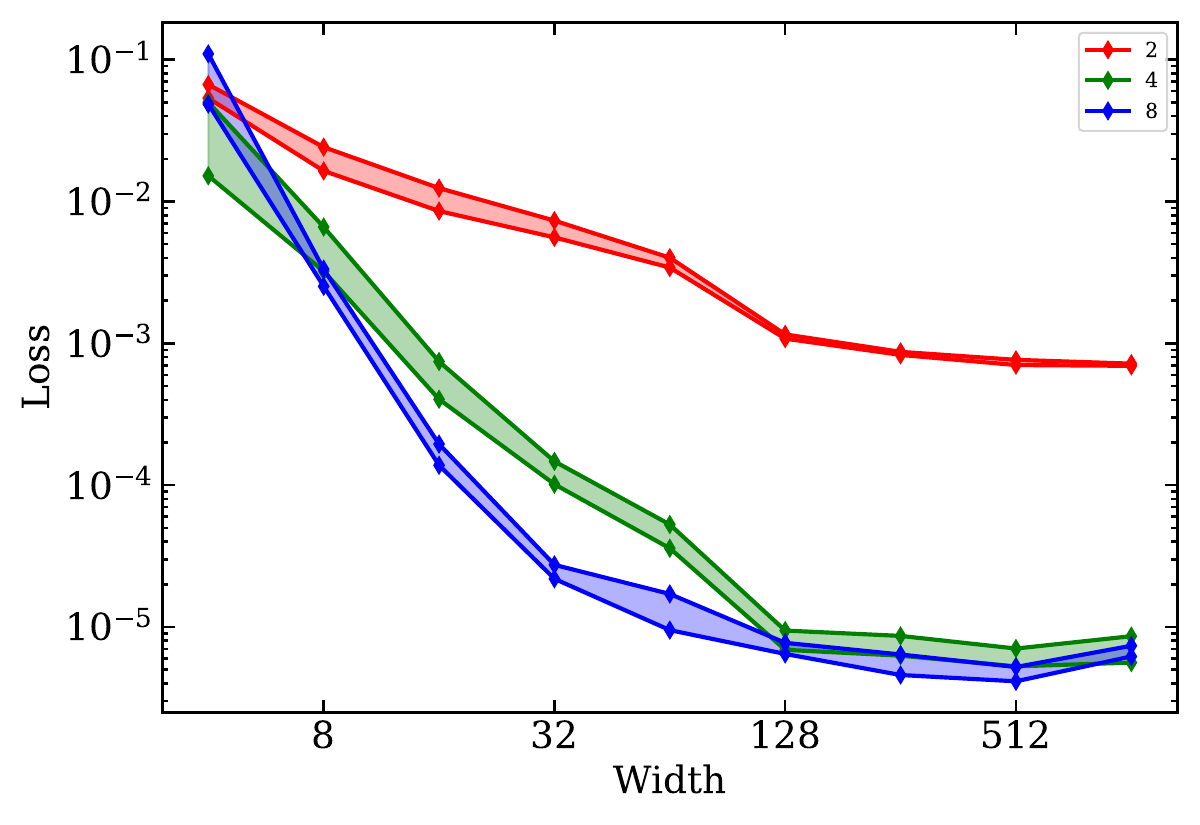}
    \caption{Mean square error loss versus width of the neural network. Different colored lines denote different depths (red depth 2, green depth 4, and blue depth 8). Ten different initializations were trained for each width and depth configuration. The bottom line is the minimum test loss of the best performing network and the top line is the minimum test loss for the fifth best performing network. Four layers are necessary and sufficient to achieve optimal loss, while increasing the width of the network also increases its accuracy.}
    \label{fig:losswidth}
\end{figure}

Figure \ref{fig:losswidth} shows the mean squared error loss of the last epoch as a function of network width with lines corresponding to different depths. It is clear that, while 2 hidden layers are not sufficient, 4 or 8 hidden layers (at large widths) drive the loss down to the floor of single precision floats. 

We can use the mean squared error as a good indication to evaluate how a neural network is performing and to compare the performance of two networks attempting to perform the same task. However, in order to measure if the neural network is achieving the required $<1\%$ accuracy, we calculate the residual in each prediction. For a given flux in the pulse profile calculated using ray tracing, $F_\text{RT}(\varphi),$ and using the neural network, $F_\text{NN} (\varphi),$ we define the residual as
\begin{equation}
    r(\varphi) \equiv \frac{F_\text{NN}(\varphi) - F_\text{RT}(\varphi)}{\langle F_\text{NN} \rangle}, \label{eqn:residual}
\end{equation}
where $\langle F_\text{NN} \rangle$ is the average flux for the whole pulse profile from the ray tracing algorithm.  

\begin{figure}[t]
    \centering
    \includegraphics[width=0.45\textwidth]{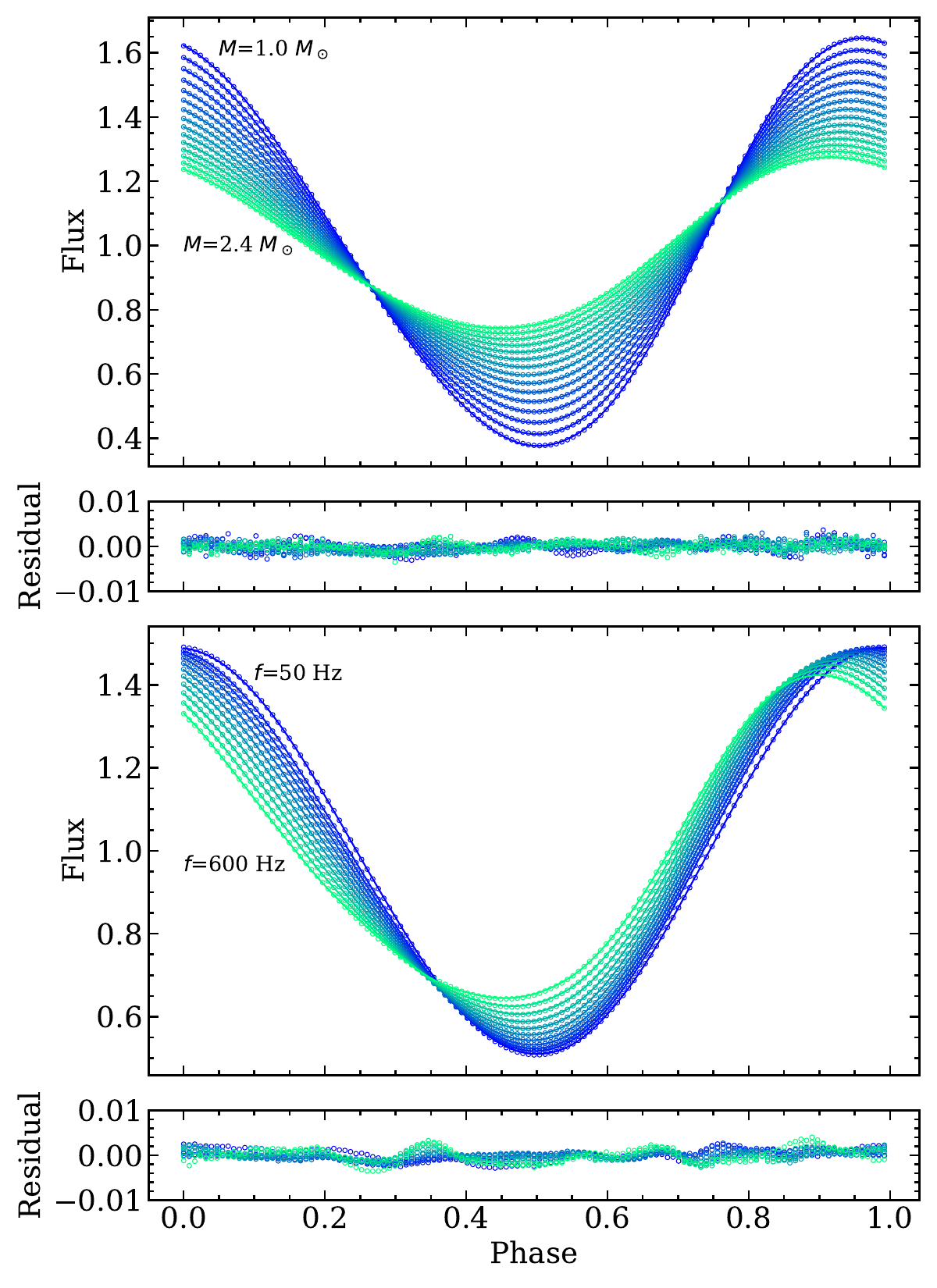}
    \caption{{\em (Top)\/} Pulse profiles from a neutron star with a mass of 1.5 $M_\odot$, a radius of 13.2 km, and a single circular hotspot with a half opening angle of 10$^\circ$ located at a colatitude of 22.5$^\circ$ from the axis of rotation. The observer is set at 75$^\circ$ from the axis of rotation. The various curves correspond to spin frequencies from 50 Hz to 600 Hz in steps of 50 Hz. Dots represent data points calculated by the ray tracing algorithm, while the solid lines are the pulse profiles generated using a trained neural network. The fractional residuals are shown in the small sub-plot and stay at a level $<0.05\%$. {\em (Bottom)\/} Same as above but for a fixed spin frequency of 300 Hz and different neutron star masses from 1~$M_\odot$ to 2.4~$M_\odot$ in steps of 0.1 $M_\odot$.}
    \label{fig:OmRTvsNN}
\end{figure}

\begin{figure*}[t]
    \centering
    \includegraphics[width=1.0\textwidth]{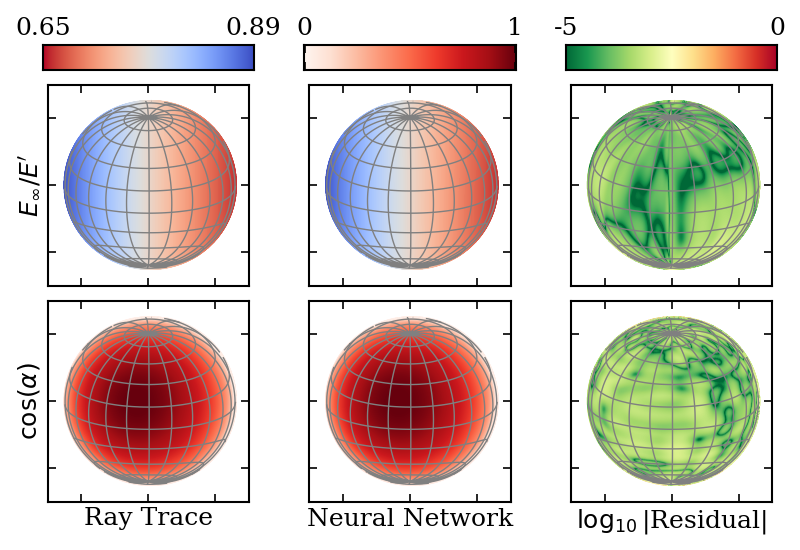}
    \caption{{\em (Top)\/} The frequency shift $E_\infty/E^\prime$ experienced by photons originating from different locations on the neutron-star surface, as calculated using the ray tracing algorithm (left) and the neural network (middle). The rightmost panel shows the absolute residual between the two. {\em (Bottom)\/} Same as the top panels but for $\cos(\alpha)$. The residuals are well below 1\% except for a very small region near the edge of the neutron-star surface.}
    \label{fig:RRTvsNN}
\end{figure*}

Figures \ref{fig:OmRTvsNN} and Figure \ref{fig:RRTvsNN} compare results from the ray tracing and trained neural networks and show the corresponding residuals. The panels with the residuals in the figures show that the neural network has indeed achieved an accuracy well within 1\% of the ray tracing output. The fact that the neural network can smoothly transition between the ray tracing points shows that the network generalized and can output fluxes within the distribution of data points trained on but not contained in the training data without introducing spurious nonphysical fluctuations. This figure also demonstrates that the neural network did not overfit the training data: as expected, the residuals are of order $\lesssim \sqrt{\text{MSELoss}}$.

\begin{figure}[t]
    \centering
    \includegraphics[width=0.45\textwidth]{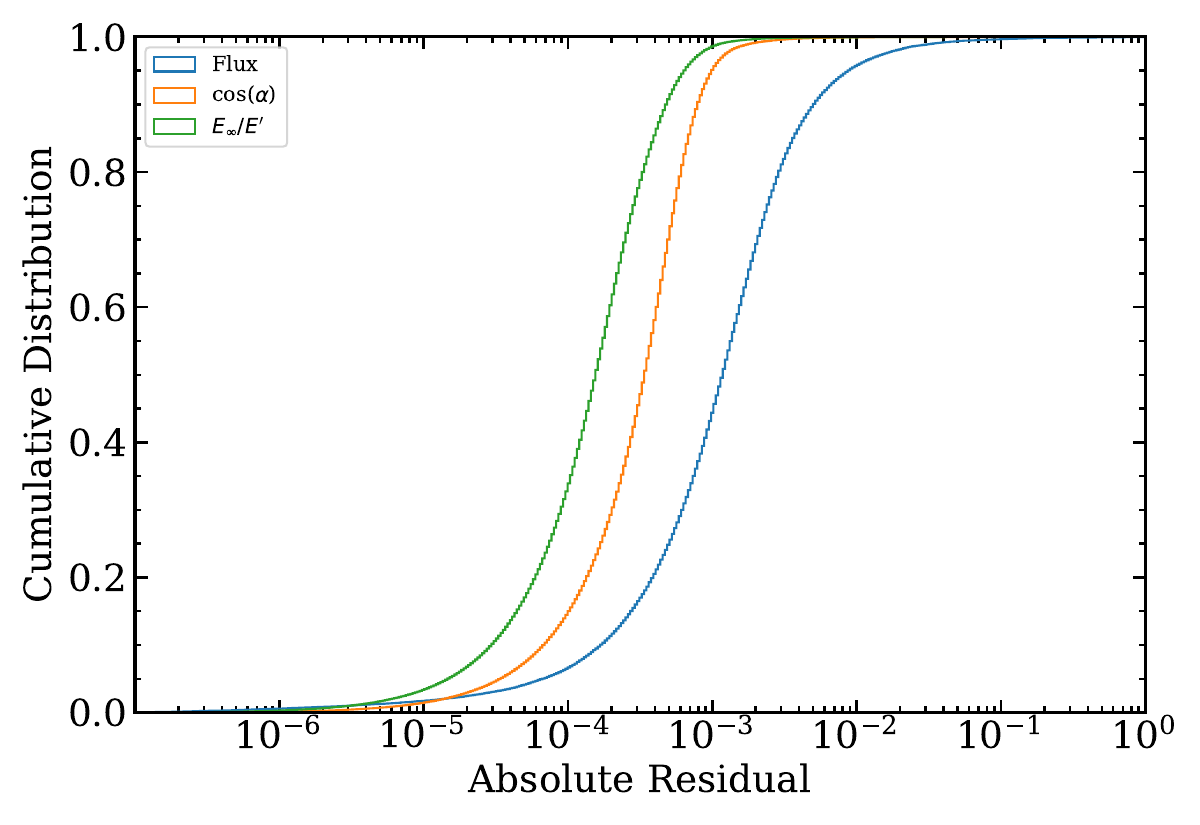}
    \caption{Cumulative distribution functions of the absolute residuals when comparing the outputs of the neural networks against the validation dataset, which does not overlap with the test or train datasets. Approximately 96\% of the absolute residuals for the flux, 99.97\% for the $\cos(\alpha)$, and 99.98\% for the $E_\infty/E^\prime$ networks fall below our target 1\% error.}
    \label{fig:valcdf}
\end{figure}

Finally, we calculated 1024 additional pulse profiles with parameters that have never been seen by the neural network and did not influence our choice of the best neural network for speed or accuracy. With this dataset as our validation dataset, we calculated the residual at each point. The cumulative distribution functions of the absolute residual for the three networks are plotted in Figure~\ref{fig:valcdf}. From this plot, we can see that 96\% of the outputs from the flux neural network, 99.97\% of the outputs from the $\cos(\alpha)$ neural network, and 99.98\% of the outputs from the $E_\infty/E_\infty$ neural network fall well within 1\%, which has been our target accuracy. 

Investigating the outputs that do not fall within the $1\%$ target,  we identified two factors that lead to the inaccuracies: (1) for some configurations, the neutron star was under resolved in the ray tracing code leading to noise in the pulse profile; (2) when the emission regions is on the edge of the visible surface, the resulting flux is very small, leading to numerical errors. From the density of data in the parameter space, we conclude that the neural networks were able to smooth out this noise and not over fit to the training data.  

\subsection{Timing} \label{subsec:time}

The primary reason for using machine learning and, therefore, a neural network to calculate the flux from a pulsar is computational efficiency. Building the ray-tracing dataset of pulse profiles on which this neural network is trained took $\sim$ 1.49 million core hours across 738 jobs using full nodes (24 cores) of the Hive cluster at the Georgia Institute of Technology Partnership for an Advanced Computing Environment (PACE). This comes out to an average of 87 minutes per pulse profile and 41 seconds per data point.  

The real benefit of the machine learning algorithm becomes apparent by the fact that the entirety of this dataset can be recreated with the neural network in 26 seconds (subtracting time to move the inputs onto the GPU), averaging 180 nanoseconds per data point for the network with depth 4 and width 1024. The timing of evaluations of the neural networks was done on a MacBook Pro with an Apple Silicon M3 Max SOC and 40 core GPU. Figure~\ref{fig:timing} shows how this timing changes with the different hyperparameters. This 9 orders-of-magnitude speed up results from the fact that the calculation is simplified down to a few matrix multiplications and calls to an activation function. With frameworks like PyTorch, these calculations can be trivially performed on a GPU in a massively parallel fashion.

\subsection{Optimal Network Architecture} \label{subsec:optimal}

\begin{figure}[t]
    \centering
    \includegraphics[width=0.45\textwidth]{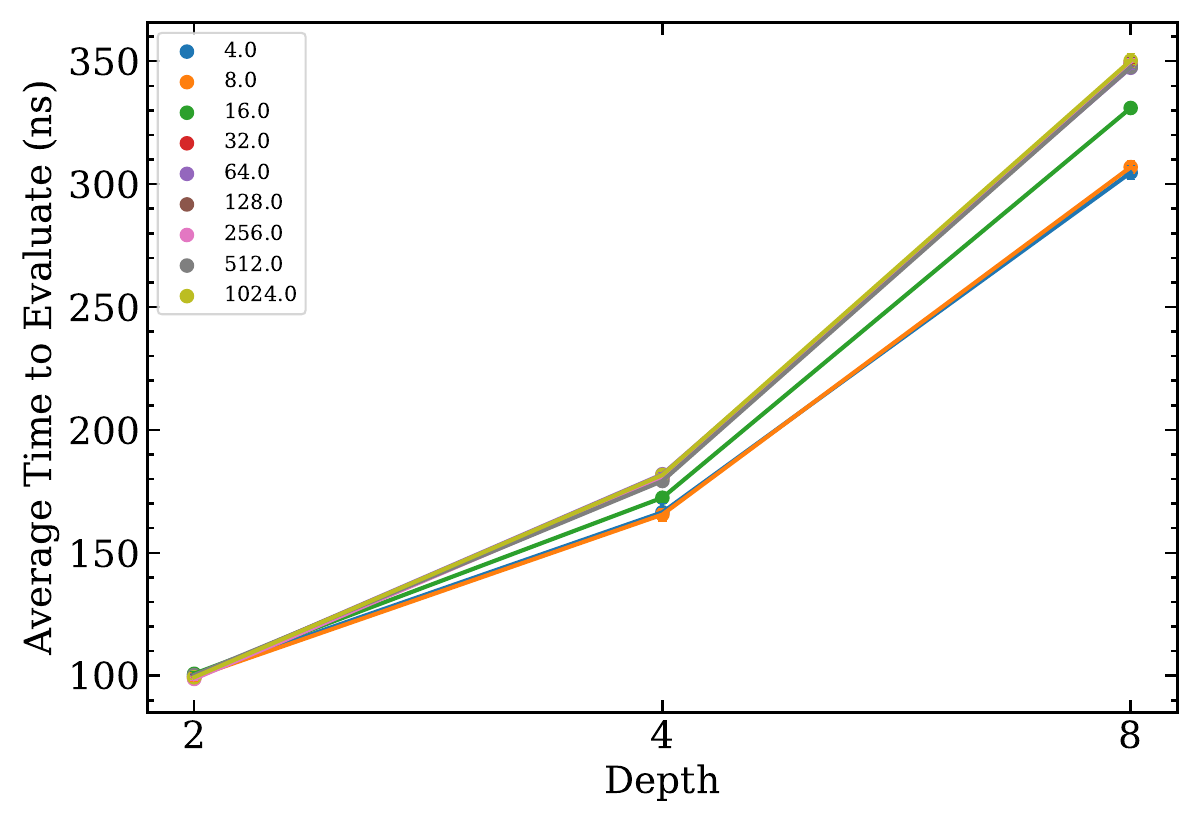}
    \caption{Average flux evaluation time (in nanoseconds) for one configuration as a function of the depth of the neural network, for different widths. The evaluation time depends primarily on the depth of the network, because increasing the width can be compensated by increasing the parallelization of the evaluation.}
    \label{fig:timing}
\end{figure}

Figure~\ref{fig:losswidth} shows that increasing the depth of the neural network from 2 to 4 improves its accuracy, but further increasing it to 8 does not lead to any substantial improvement. Similarly, Figure~\ref{fig:timing} demonstrates that the evaluation time for the networks depends primarily on their depth. Combining these two pieces of information, we conclude that a depth of 4 optimizes the accuracy of the network while not requiring unnecessarily long evaluation times.

Similarly, Figure~\ref{fig:losswidth} shows that increasing the width of a network of depth 4 to beyond 128 neurons per layer does not improve the accuracy of the evaluation. As a result, we  settle on a network of depth 4 and width of 128 as the optimal network architecture.

\section{Pulse Profiles of Arbitrary Geometries} \label{sec:geom}

Using the three neural networks for flux, frequency shift $E_\infty/E^\prime$, and beaming angle $\cos(\alpha)$, we can construct pulse profiles at different energies for any emission geometry we desire. In this section, we will demonstrate the process for this calculation using a configuration with a neutron star of radius 11.5 km, mass 1.8 $M_\odot$, frequency 350 Hz, and observing angle of $74.375^\circ$. 

\begin{figure}[t]
    \centering
    \includegraphics[width=0.49\textwidth]{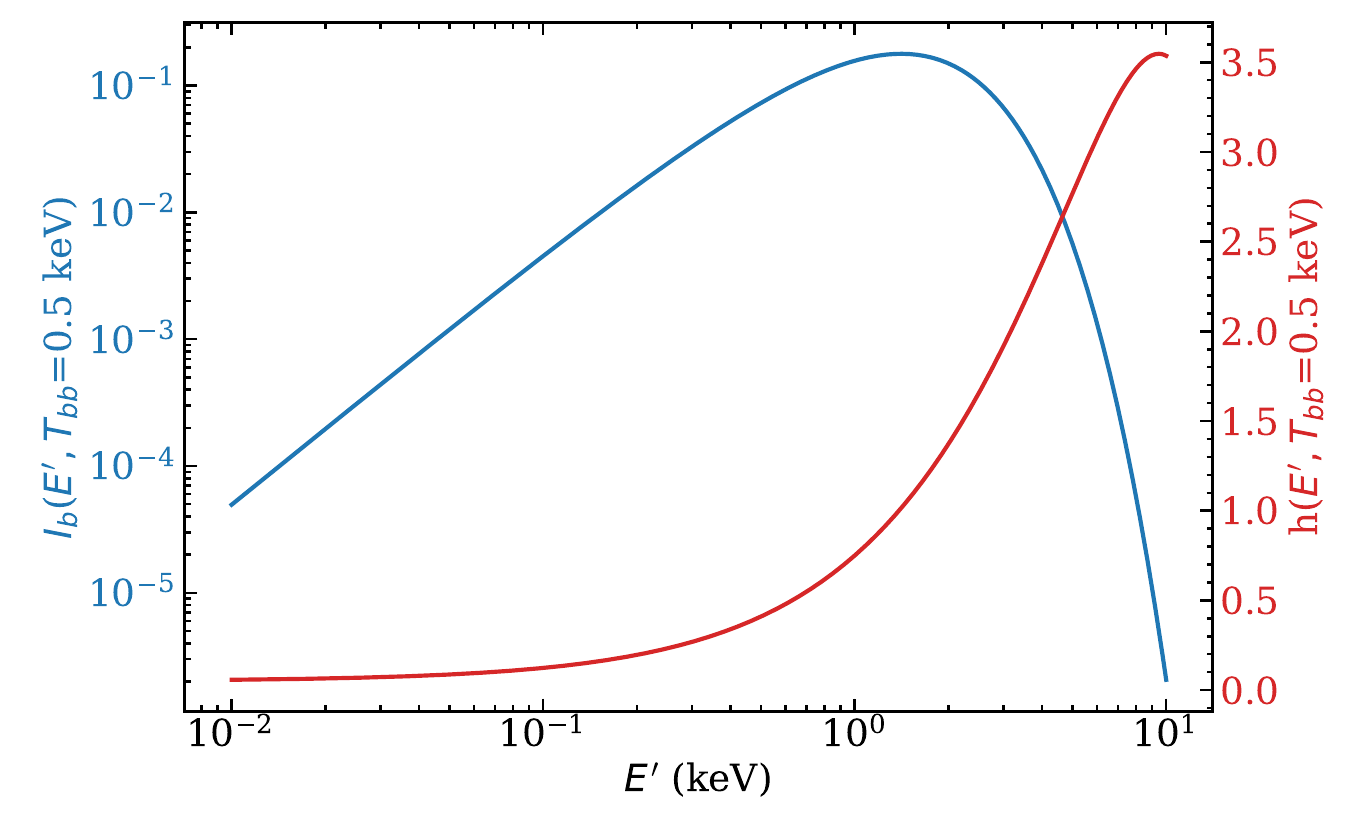}
    \caption{The blackbody $I_{b}(E^\prime,T_{bb})$ and beaming functions $h(E^\prime,T_{bb})$ we use for the example discussed in \S\ref{sec:geom}, plotted as a function of photon energy on the surface of the neutron star, $E^\prime$. Here $T_{bb}=0.5$~keV.}
    \label{fig:hIb}
\end{figure}

\begin{figure*}[t]
    \centering
    \includegraphics[width=1.0\textwidth]{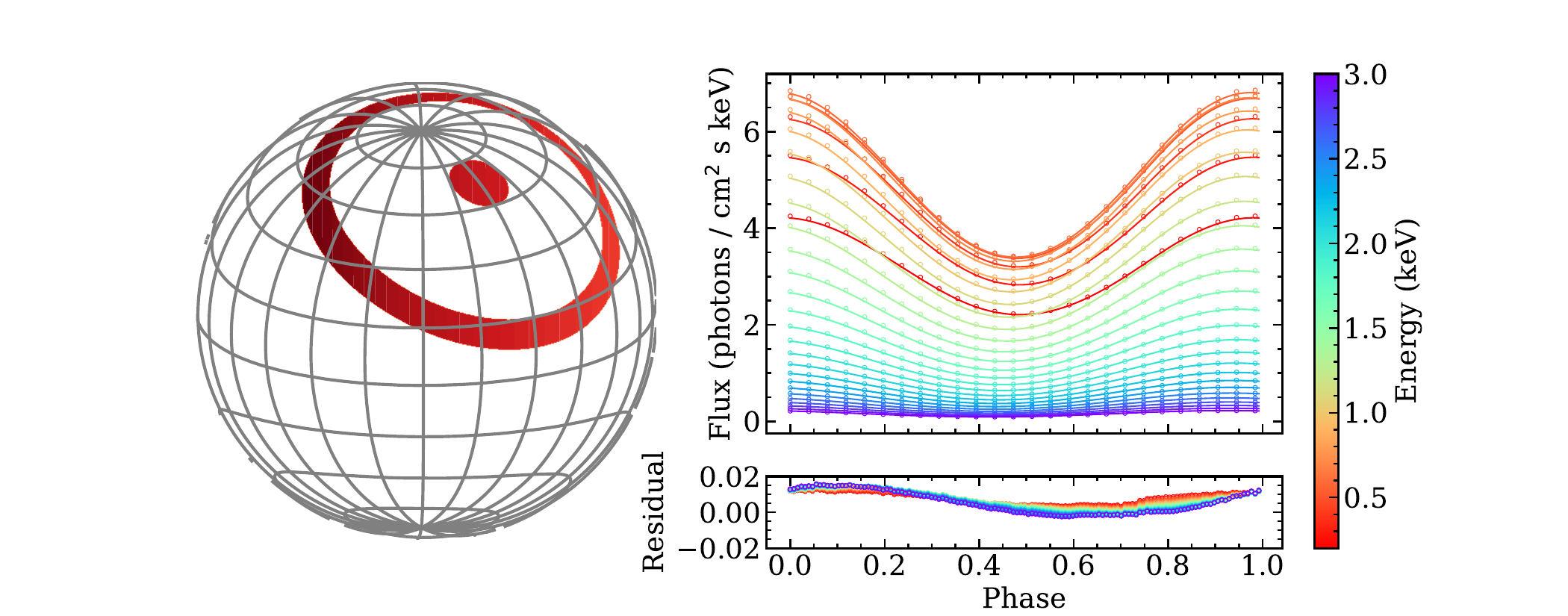}
    \caption{{\em (Left)\/} Projection onto the image plane of the surface of a pulsar with a $10^\circ$ hotspot and a concentric ring extending between $50^\circ$ to $60^\circ$ from the center of the hotspot. This pulsar has a radius of 11.5 km, mass of 1.8 $M_\odot$, spin frequency of 350 Hz, the center of the hotspot is $30^\circ$ from the axis of rotation, the observer is at $74.375^\circ$ from the axis of rotation, and the surface emission is that of a deep-heated atmosphere with a blackbody temperature of 0.5~keV. {\em (Right)} The observed flux as a function of phase, as calculated with the ray tracing algorithm (dots) and with the neural networks (lines). The colors correspond to different photon energies of photons at infinity.}
    \label{fig:spotring}
\end{figure*}

For the surface emission, we use a hotspot of half opening angle of $10^\circ$ centered on the ``polar axis'' of the star and a concentric ring extending from $50^\circ$ to $60^\circ$ from the center of the spot. For the spectrum emerging from the spot and ring, we use a blackbody function with a temperature of $T_{bb}=0.5$~keV and choose a beaming function $h(E^\prime,T_{bb})$ corresponding to a deep heated atmosphere (see \citealt{TongBeam}), i.e., we use
\begin{equation}
    I(E^\prime,T_{bb},\alpha) = I_b(E^\prime,T_{bb}) \frac{1+h(E^\prime,T_{bb}) \cos(\alpha)}{1+(2/3)h(E^\prime,T_{bb})}, 
\end{equation}
where the blackbody function is given by
\begin{equation}
    I_b(E^\prime,T_{bb}) = \frac{(E^\prime)^3 }{e^{E^\prime/T_{bb}} -1},
\end{equation}
and the beaming function is
\begin{equation}
    h(E^\prime,T_{bb}) = 0.05 + 0.3695 \left( \frac{E^\prime}{T_{bb}} \right) - 0.00976 \left( \frac{E^\prime}{T_{bb}} \right)^2.
\end{equation}
Figure~\ref{fig:hIb} shows the energy dependence of the blackbody and beaming functions assumed here.

We can decompose the complex emitting region (i.e., the spot+ring) into infinitesimal surface elements and then integrate the flux $F(\vec{\lambda}; \theta,\varphi)$ that we observe at infinity from each of those infinitesimal surfaces to calculate the total observed flux. Formally, we write 
\begin{eqnarray}
    F(\vec{\lambda}; E, \varphi) &=& \int_0^{2\pi} \int_0^\pi d\theta d\phi F(\vec{\lambda}; \varphi+\phi)\nonumber\\ 
    && I(E/g(\vec{\lambda};\theta,\varphi+\phi), T_{bb}, \alpha(\vec{\lambda};\varphi+\phi))\;.
    \label{eq:integral}
\end{eqnarray}
Here, $\vec\lambda=(R,M,\Omega, \theta_i, \varphi+\Delta \varphi_i)$ is the vector of parameters that describes the particular configuration, and $F(\vec{\lambda}; \varphi)$, $g(\vec{\lambda};\theta,\varphi)$, and $\alpha(\vec{\lambda};\varphi)$ are the outputs of the flux, energy-shift, and beaming angle neural networks.

It is important to emphasize here that, even though the calculation of the observed flux from a complex emission geometry requires the evaluate of a two-dimensional integral when using our neural-network algorithm, it still remains computationally less expensive than a traditional ray-tracing approach, for two reasons. First, the evaluation of the flux from each infinitesimal surface area on the stellar surface does not require integration of geodesics but is evaluated efficiently using the neural networks. Second, the integral in expression~(\ref{eq:integral}) is performed over the neutron-star surface and not over the observer's image plane, as is the case for traditional ray-tracing algorithms with spacetimes that are not spherically symmetric. This allows us to specify an integration grid on the stellar surface that is independent of the spin phase and can efficiently follow the contours of surface emission, neither of which are possible to do on the image plane.

Figure \ref{fig:spotring} shows a comparison between the pulse profiles at different photon energy calculated using the ray-tracing algorithm and the neural network. In order to perform the surface integral using the neural network, we use 720,000 points with separations in $\theta$ and $\phi$ of $\Delta\theta=\Delta\phi=0.1^\circ$. Even for this complex configuration, the calculation using the neural networks is accurate to within our target 1\% error. More importantly, even though the ray tracing algorithm took 150 minutes to complete the evaluation, the neural network took only 11.13 seconds, i.e., an acceleration by a factor of $\sim 800$. 

\section{Conclusion} \label{sec:concl}

In this work, we presented a machine learning algorithm to calculate pulse profiles from thermally emitting neutron stars. The algorithm incorporates three neural networks that allow the determination of the observed flux, the energy shift, and the beaming angle of photons that emerge from any location on the neutron-star surface. The evaluations are accurate to within 1\% of those generated with a ray tracing algorithm that includes the Hartle-Thorne metric, Doppler boosting, time delays, as well as the oblate shapes of the neutron star surfaces.

\begin{figure}[t]
    \centering
    \includegraphics[width=0.45\textwidth]{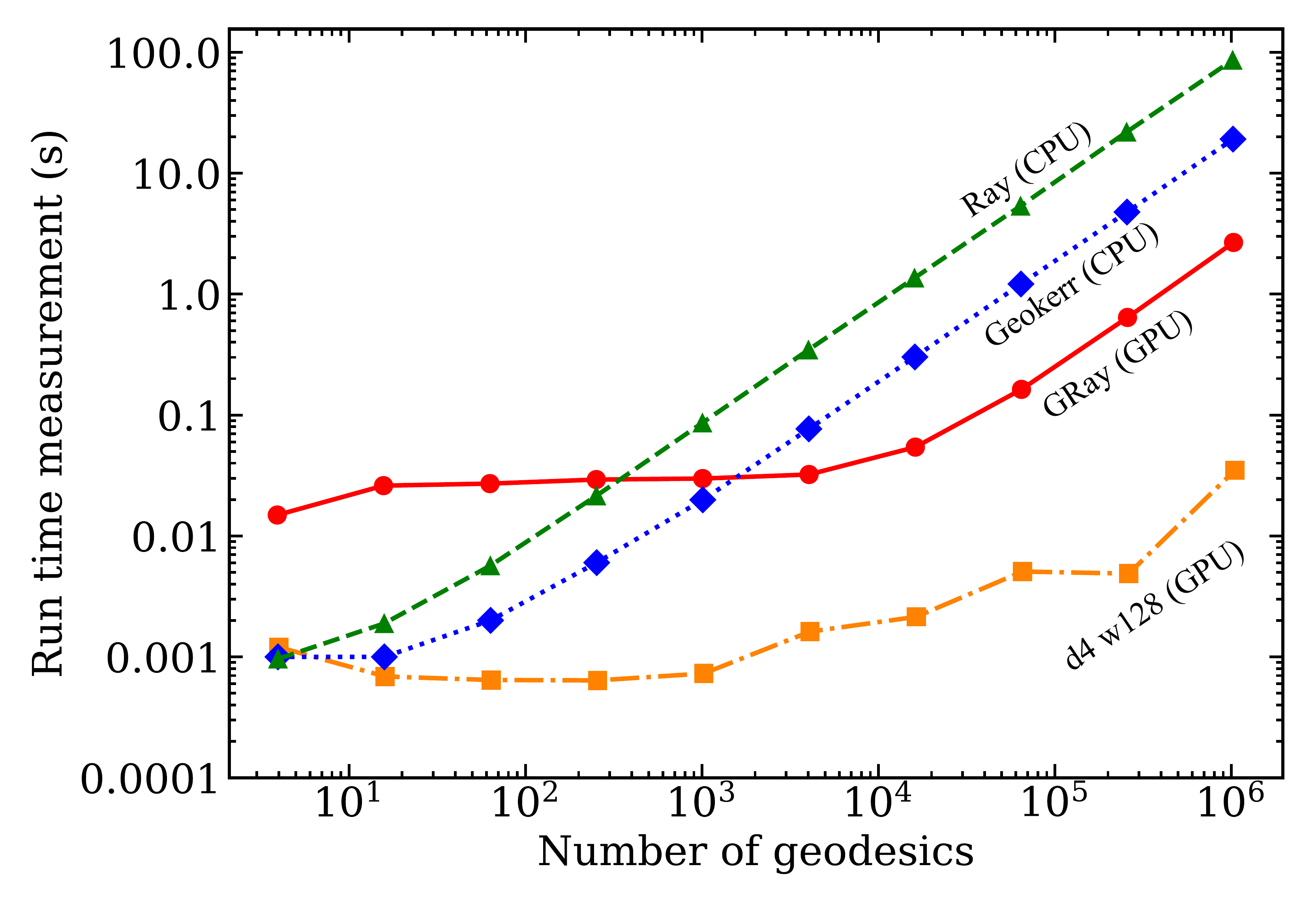}
    \caption{Run time as a function of the number to geodesics calculated, for different approaches to ray tracing. \texttt{Ray} is a CPU-based numerical integrator, \texttt{Geokerr} is a semi-analytic integrator, and \texttt{GRay} is GPU-based. The orange curve shows the run time for the neural network of depth 4 width 128 demonstrating $\sim 2$ order of magnitude acceleration even with respect to the GPU-based algorithm. }
    \label{fig:GRayComp}
\end{figure}

Our motivation to create this machine learning algorithm has been to generate  pulse profiles with a fast and efficient method that allows large-dimensional parameter inference from data. Figure~\ref{fig:GRayComp} compares the performance of the machine learning algorithm with those of traditional ray-tracing approaches, including the numerical CPU-based \texttt{Ray} algorithm~\citep{Baubock2012}, the semi-analytical \texttt{GeoKerr} algorithm~\citep{Dexter2009}, and the GPU-based numerical algorithm \texttt{GRay}~\citep{GRay}. 

For this comparison, the ``number of geodesics'' for the machine-learning algorithm measures the batch size used. This is appropriate since, for a complex surface emission geometry, a number of evaluations will be needed to calculate the integral in expression~(\ref{eq:integral}). The run time, which also includes the time to move the input data onto the GPU, has been averaged over the evaluation of 100 batches. For this comparison, the machine learning algorithm is $\sim 2$ orders of magnitude faster than the GPU-accelerated one and four orders of magnitude faster than CPU-based algorithm. Such an improvement in efficiency is sufficient to enable a similar increase to the number of MCMC samples used when fitting NICER data and alleviate the challenges introduced by this computational bottleneck.  

\section*{Acknowledgments}
We would like to thank Feryal \"Ozel and the Xtreme Astrophysics group at Georgia Tech for their comments and discussions that contributed to this manuscript. 

\bibliography{export-bibtex}{}
\bibliographystyle{aasjournal}

\end{document}